# Suppressing the cellular breakdown in Silicon supersaturated with Titanium

**Fang Liu[1,2], S Prucnal[1], R Hübner[1], Ye Yuan[1,2], W Skorupa[1],**

**M Helm[1,2] and Shengqiang Zhou[1]**

[1]Helmholtz-Zentrum Dresden-Rossendorf, Institute of Ion Beam Physics

and Materials Research, Bautzner Landstr. 400, 01328, Dresden, Germany

[2]Technische Universität Dresden, Dresden, 01062, Germany

E-mail: f.liu@hzdr.de

**Abstract.** Hyper doping Si with up to 6 at.% Ti in solid solution was performed by ion implantation followed by pulsed laser annealing and flash lamp annealing. In both cases, the implanted Si layer can be well recrystallized by liquid phase epitaxy and solid phase epitaxy, respectively. Cross-sectional transmission electron microscopy of Ti-implanted Si after liquid phase epitaxy shows the so-called growth interface breakdown or cellular breakdown owing to the occurrence of constitutional supercooling in the melt. The appearance of cellular breakdown prevents further recrystallization. However, the out-diffusion and cellular breakdown can be effectively suppressed by solid phase epitaxy during flash lamp annealing due to the high velocity of amorphous-crystalline interface and the low diffusion velocity for Ti in the solid phase.

**Keywords:** Ion implantation, Solid phase epitaxy, Liquid phase epitaxy, Si, Cellular breakdown

## 1. Introduction

Impurities play an important role in determining the electrical and optical properties of semiconductors. Recently it was demonstrated that deep level impurities can create an impurity band inside the gap of semiconductors if the doping concentration is high enough, e.g. for titanium (Ti) or chalcogens in Si [1-3]. The insertion of an impurity band can enhance the near infrared light absorption and leads to applications in the so-called intermediate band solar cell [2, 4, 5] or as infrared photodetectors [2, 6]. To form such an impurity band instead of isolated levels, one needs to have doping concentrations far exceeding the Mott transition limit, which is ~$5\times10^{19}$ cm$^{-3}$ for Ti [7]. Unfortunately, most deep level impurities (Ti, S and Se) have relatively low solubility in silicon [8-10]. As an industry-standard method, ion implantation works in non-thermal-equilibrium regime and can overcome the thermal solubility limit. However, high-fluence ion implantation damages the crystalline order. Pulsed laser annealing (PLA) has been widely used to anneal Ti and chalcogen-ion-implanted Si [4, 11-13]. Using PLA, the surface layer can be molten in the nanosecond range and recrystallized via liquid-phase epitaxy [14, 15]. In order to attain supersaturation, the solidification rate must be larger than the diffusive speed of the impurities [16-18]. However, the challenge of PLA is to avoid dopant segregation and cellular breakdown at high doping concentration, preventing supersaturation. Regarding Ti-implanted Si, Pastor *et al.* reported that Ti ions massively occupy interstitial sites in the host semiconductor in highly Ti-implanted Si layers followed by PLA [19]. Recently, Mathews *et al.* pointed out that the intrinsic limit of the Ti concentration in Si is around $1…2\times10^{21}$ cm$^{-3}$ (2-4 at. %) [20]. This is attributed to the fact that during pulsed laser annealing and rapid solidification the impurities diffuse to the surface of the implanted layer and form a cellular breakdown microstructure, preventing further recrystallization [20].

In our work, single crystalline Si wafers have been implanted with Ti concentration above the concentration limit ($1…2\times10^{21}$ cm$^{-3}$) mentioned in ref. 20. The implanted layers were annealed by PLA and flash lamp annealing (FLA). Different from PLA, FLA in millisecond range can render the sample just slightly below the melting temperature leading to solid-phase epitaxy [21]. Our results indicate that Ti-implanted Si with a concentration above the equilibrium solid solubility limit can be recrystallized by both PLA and FLA. In terms of preventing surface segregation, FLA is superior to PLA because the cellular breakdown can be avoided leading to a much higher incorporation concentration.

## 2. Experimental

Intrinsic (001) Si wafers with a thickness of 525 μm (R > $10^4$ Ω·cm measured at room temperature) were implanted with a Ti$^+$ fluence of $1.2 \times 10^{16}$ cm$^{-2}$ at 35 keV. After implantation, the wafers were cut into pieces of $5 \times 5$ mm$^2$ for PLA and FLA. The FLA was performed with a 3 ms light pulse in Ar ambient and the optical energy fluence was varied from 20 to 100 J/cm$^2$. The estimated peak temperatures at the sample surface were below 1300 °C. For comparison, some samples were annealed using a Coherent XeCl excimer laser (λ=308 nm) at 0.6-1.0 J/cm$^2$ with a single pulse of 28 ns in air atmosphere. The structural properties were investigated by Rutherford backscattering spectrometry/channeling (RBS/C) and Raman spectroscopy. For RBS measurements, 1.7 MeV He$^+$ ions were used at a backscattering angle of 170°. The RBS spectra were measured along random and channeling geometry. The Raman scattering spectra were recorded in the backscattering geometry using a 532 nm Nd:YAG laser with a charge coupled device camera cooled with liquid nitrogen. To locally analyze the microstructure of the Ti-implanted and subsequently annealed Si films, cross-sectional transmission electron microscopy (TEM) images were taken using an image-corrected FEI Titan 80-300 microscope operated at 300 kV. Element mapping based on energy-dispersive X-ray spectroscopy (EDXS) was performed using a FEI Talos F200X microscope operated at 200 kV and equipped with a Super-X EDXS detector system. In this manuscript, we only show the results annealed under the optimized conditions, i.e. obtaining the best recrystallization of Si, which are 0.8 J/cm$^2$ for PLA and 55.5 J/cm$^2$ for FLA, respectively.

## 3. Results

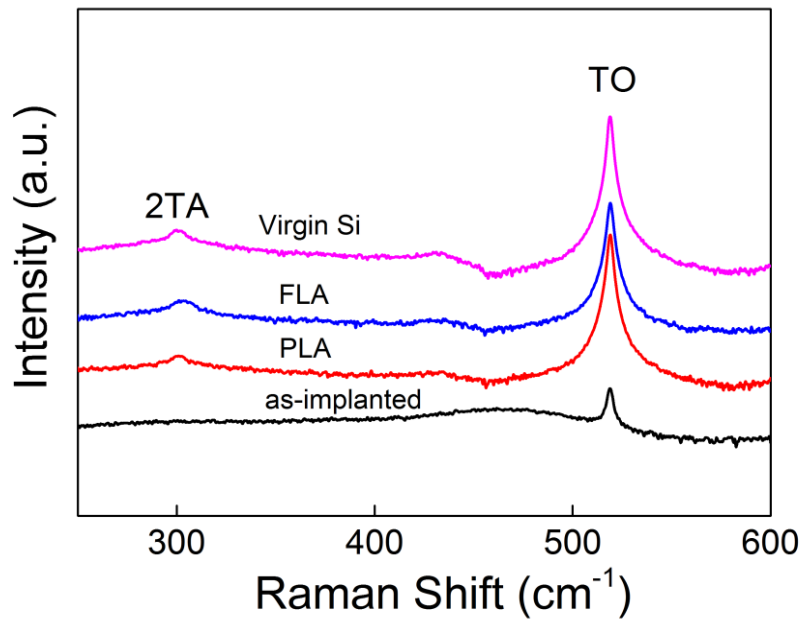

Figure 1. Raman scattering spectra of Ti-implanted Si samples with a fluence of $1.2 \times 10^{16}$ cm$^{-2}$ and subsequently annealed by PLA and FLA. For comparison, the results of as-implanted and virgin Si are also shown.

In figure 1, we show the Raman spectra of Ti-implanted Si samples with a fluence of $1.2 \times 10^{16}$ cm$^{-2}$ before and after PLA or FLA. The Raman spectrum of the single crystalline Si substrate is depicted for reference. The spectrum of the as-implanted sample exhibits a weak band at around 460 cm$^{-1}$, which is a typical feature of the amorphous silicon formed by the ion implantation process [22]. In addition, the as-implanted sample also shows a weak peak at around 520 cm$^{-1}$ corresponding to the TO mode of single crystalline Si substrate beneath the implanted amorphous layer. After PLA or FLA, the weak band at 460 cm$^{-1}$ which is related to the amorphous Si layer vanishes. The TO mode at 520 cm$^{-1}$ appears as a sharp peak with similar intensity and linewidth of that from virgin single crystalline Si. The peak at 303 cm$^{-1}$ corresponds to the two transverse acoustic (2TA) phonon mode of Si, which indicates that the crystalline order can be well recovered.

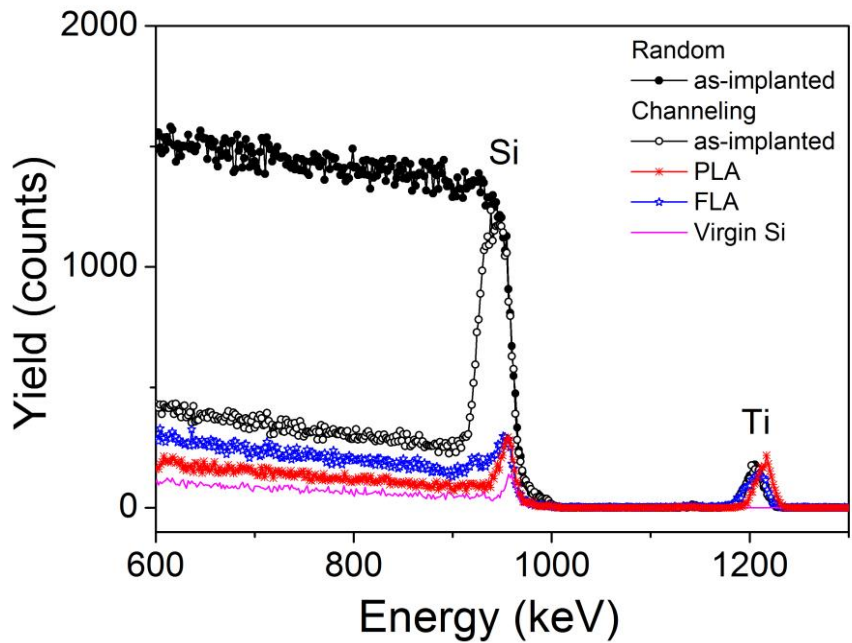

Figure 2. RBS spectra of Ti-implanted Si samples with a fluence of $1.2 \times 10^{16}$ cm$^{-2}$ at random and channeling configuration after annealing ( 0.8 J/cm$^2$ for PLA and 55.5 J/cm$^2$ for FLA). The spectra of as-implanted and virgin Si are also included for comparison.

To study in more detail the crystallization process, RBS/Channeling measurements were performed on the samples before and after annealing. Figure 2 shows the RBS spectra of the implanted samples in both random and channeling configurations. As we can see, the as-implanted sample displays a broad damage peak at around 940 keV, pointing out that a highly disordered or amorphous Si layer has been formed during ion implantation. However, after annealing (by both PLA and FLA), the channeling spectra of the annealed samples with a very low backscattering yield are similar to that of virgin Si, indicating that the crystal order can be quite well reconstructed with considerable lattice quality. This is in a good agreement with the results obtained by Raman spectroscopy.

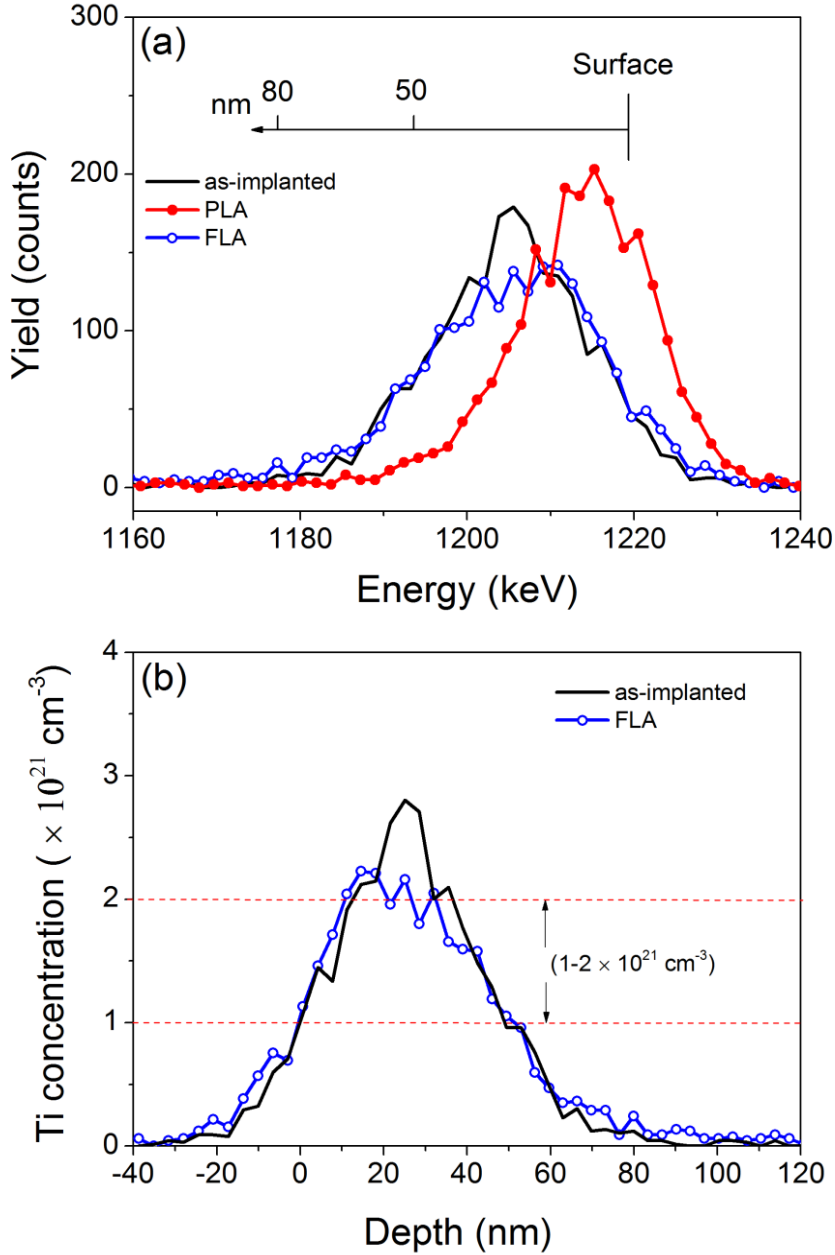

Figure 3. (a) RBS spectra of Ti-implanted Si samples with a fluence of $1.2 \times 10^{16}$ cm$^{-2}$ before and after annealing. A depth scale is indicated in the figure. (b) The depth profile of Ti in Si implanted with a fluence of $1.2 \times 10^{16}$ cm$^{-2}$ before and after FLA, which is calculated by using RUMP[23]. The Mott transition limit is around $5 \times 10^{19}$ cm$^{-3}$ for Ti in Si (Ref. 7). The range between the two dash lines in the figure is the Ti concentration limit obtained by liquid phase epitaxy[20].

In addition to prove the crystalline nature of implanted Si by FLA or PLA, RBS can be used to determine the depth distribution of Ti. We show in figure 3(a) the magnified RBS spectra corresponding to the as-implanted sample and the annealed samples. Compare with the as-implanted sample, the Ti-related peak of the sample annealed by PLA shifts to higher energy, which indicates that Ti atoms diffuse towards the surface. In contrast, there is no significant difference before and after FLA. Figure 3(b) shows the Ti concentration-depth profile. The Ti concentration in FLA sample is well above Mott limit, which is also above the concentration limit (range between two dash lines) mentioned in ref. 20. From these results, it is concluded that Ti impurities diffuse to the surface of the implanted layer during the PLA while there is no such diffusion during FLA treatment. The limit of impurity incorporated into the single crystalline Si matrix can be enhanced by FLA.

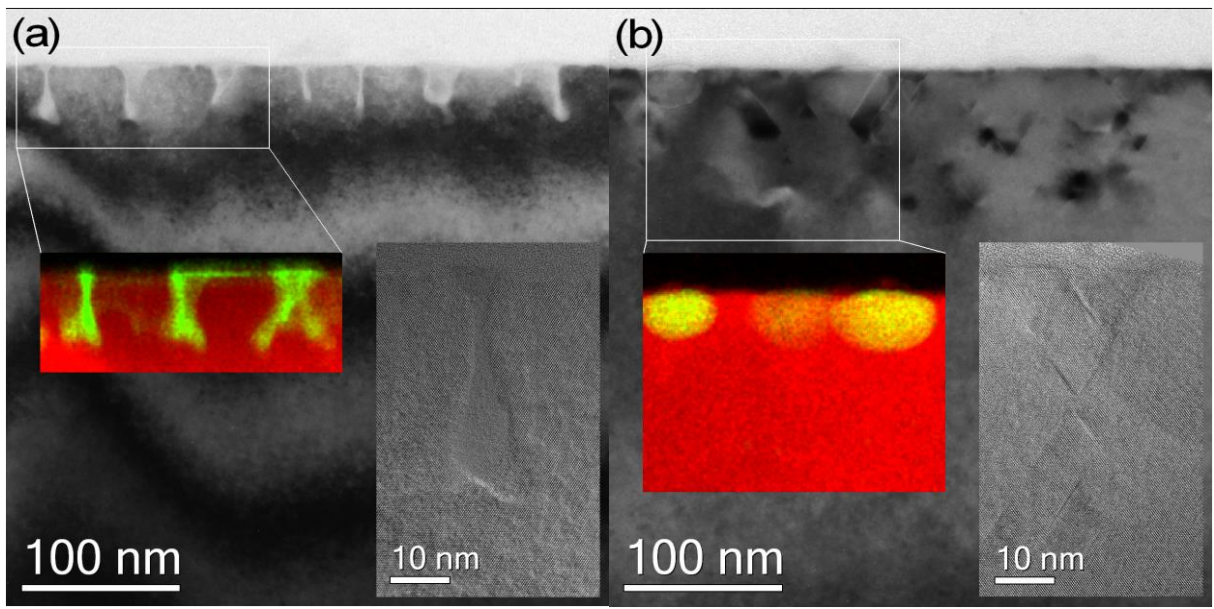

Figure 4. Cross-sectional bright-field TEM and high-resolution TEM micrographs (right insets) together with EDXS mappings (Ti: green, Si: red) of the indicated rectangular regions of Ti-implanted Si samples with a fluence of $1.2 \times 10^{16}$ cm$^{-2}$ (a) after PLA: 0.8 J/cm$^2$ and (b) after FLA: 55.5 J/cm$^2$.

Figure 4(a) shows a cross-sectional bright-field TEM micrograph of Ti-implanted Si after PLA. According to comparable diffraction contrast extending from the Si substrate to the sample surface, single crystalline regrowth of the Ti-implanted layer occurs during pulsed laser annealing. There are, however, amorphous "walls" with a lateral separation of about 40 nm extending from the sample surface almost perpendicularly to a depth of up to 40 nm. HRTEM imaging (right inset of figure 4(a)) confirms the amorphous microstructure of these walls and the defect-free single-crystalline regrowth alongside. According to EDXS mapping (left inset of figure 4(a)), titanium is mainly observed within the walls and partially at the sample surface. The measured Ti signal in the neighboring areas can be explained by a projection effect of the 2d mapping probing not only the regrown Si areas but also a Ti-containing wall lying at least partially in the plane of the TEM lamella. The TEM results indicate that a cell structure has formed during the rapid solidification process after PLA. It is the so-called growth interface breakdown or cellular breakdown due to the occurrence of constitutional supercooling in the melt [24-27]. At high enough impurity concentrations, a morphological instability occurs at the liquid-solid interface, resulting in the lateral segregation of Ti impurities and leading to a cellular solidification microstructure [24-27]. At low impurity concentration, the planar interface can still be kept but with surface segregation [19]. As shown in figure 4(a), the average lateral cell dimension in Ti-implanted Si after PLA is around 40 nm. The interior of each cell is defect-free silicon with Ti segregating at the cell walls. The cell size is proportional to the ratio between the impurity diffusion coefficient in the liquid silicon and the velocity of solidification [28]. The appearance of cellular breakdown prevents impurity incorporation into the single crystalline Si matrix.

In comparison, figure 4(b) shows a bright-field TEM image of Ti-implanted Si after FLA. Again, single crystalline regrowth occurs during annealing. However, defects, such as stacking faults (right inset of figure 4(b)), are incorporated into Si. The formation of such stacking faults is due to the high solidification speed during the rapid recrystallization of Si [29, 30]. Additionally, hemispherical particles beneath the sample surface are seen in figure 4(b). They are of single crystalline nature, partially misaligned compared to the Si substrate and composed of Ti and Si, as shown by EDXS mapping in the left inset of figure 4(b). Compared to PLA treatment, there is no cellular breakdown during flash lamp annealing for Si implanted with the same Ti concentration.

The impurity incorporation of Ti implanted Si was interpreted by Continuous Growth Model (CGM) for solute trapping [17],

$$k = \frac{k_e + \frac{v}{v_D}}{1 + \frac{v}{v_D}}, \tag{2}$$

where $k$ is the partition coefficient, which is the ratio of the dopant concentration in the solid and in the liquid at the interface, $k_e$ is the equilibrium partition coefficient, $v$ is the solidification velocity, and $v_D$ is the diffusive velocity.

We try to interpret how to incorporate the impurities into Si during rapid solidification. The supersaturation is due to solute trapping at the moving amorphous/crystalline interface when the solidification velocity is larger than the diffusive velocity [31, 32]. The out-diffusion of impurities happens when the diffusive velocity is larger than the solidification velocity during recrystallization. The value of $k$ is an important indicator to determine how many impurity atoms are incorporated into the semiconductor. If $k \ll 1$, the impurity atoms have more probability of redistribution and have less probability of incorporation. The redistribution means that the dopant diffuses deep into the host semiconductor or towards the surface. If $k \approx 1$ ($v$ is comparable to or larger than $v_D$), the impurity atoms have less chance of redistribution and have more probability of incorporation [32]. During liquid phase epitaxy the diffusive velocity of transition metals is typically $10^2$–$10^4$ m/s [23], and the crystallization velocity is 1-10 m/s [34]. The value of $k$ is below $10^{-2}$ for transition metals in Si [35], which is far below 1. So Ti has more chance of redistribution in Si after PLA. This can be the explanation of Ti diffusion towards the surface after annealing, which is shown in figure 3. Solute concentration profile shows a spike due to the larger $v_D$ in the liquid at the solidification front [20], which is caused by dramatically solute partitioning at the liquid-solid interface. This spike increases the chance of cellular breakdown, the morphologically instability happens at the liquid-solid interface, resulting in the lateral segregation of Ti impurities and leading to a cellular breakdown microstructure in figure 4 (a). However, the cellular breakdown cannot be observed in sample after FLA. FLA is a solid-phase process and it has a millisecond range processing time. The metastable solubility is the competition between the diffusive velocity of the impurity in the amorphous phase and the velocity of solid phase epitaxy, both of which depend on the temperature [32, 36]. The impurities can be trapped into Si in this process, due to the high velocity ($10^{-1}$–$10^{-3}$ m/s) of the amorphous-crystalline interface and the low diffusion velocity ($10^{-2}$–$10^{-4}$ m/s) for FLA (for temperatures below 1300 °C) [10, 37]. Thus, we can obtain the metastable solubility which is around one to two orders of magnitude larger than the maximum equilibrium solubility [32]. This is also why Ti redistribution can be suppressed in the FLA sample, as shown in figure 3 and figure 4(b). Burton *et al.* discussed the distribution of impurity in the crystal grown from melt [33]. In their model the impurity distribution also strongly depends on the comparison between the solidification front velocity and the impurity diffusion velocity.

## 4. Conclusions

In conclusion, we have used FLA in the millisecond time scale to incorporate titanium into the Si matrix well above the solid solubility limit. Our results indicate that a very high degree of lattice recrystallization has been attained after FLA. In terms of preventing surface segregation and cellular breakdown, FLA is superior to the nanosecond laser annealing. This approach of combining ion implantation and FLA is not limited to the case of Ti in Si only and should be generally valid for other deep level impurities, such as Au and Zn, after carefully optimizing the annealing parameters.

## Acknowledgments

We acknowledge the ion implantation group at HZDR for performing Ti implantations. Financial support from the Helmholtz-Gemeinschaft Deutscher Forschungszentren (VH-NG-713) is gratefully acknowledged. The author (FL) was supported by China Scholarship Council (File No.201307040037).